\documentclass[twocolumn,aps,prd,superscriptaddress,nofootinbib]{revtex4-2}
\usepackage{graphicx}
\usepackage{color}
\graphicspath{{figures/}{fig/}}
\usepackage{amsmath}
\usepackage{amssymb}
\usepackage{bm}
\usepackage{slashed}
\usepackage{epsfig}
\usepackage{amsfonts}
\usepackage{epstopdf}
\usepackage{hyperref}
\usepackage{bbm}
\usepackage{textcomp}
\usepackage{color}
\usepackage{booktabs}
\usepackage{ulem}
\usepackage{comment}
\usepackage{tikz-feynman}

\begin{document}
\title{Compton Scattering Total Cross Section at Next-to-Next-to-Leading Order and Resummation of Leading Logarithms}

\author{Hai Tao Li}
\email{haitao.li@sdu.edu.cn}
\affiliation{School of Physics, Shandong University, Jinan, Shandong 250100, China}
\author{Yan-Qing Ma}
\email{yqma@pku.edu.cn}
\affiliation{School of Physics, Peking University, Beijing 100871, China}
\affiliation{Center for High Energy Physics, Peking University, Beijing 100871, China}

\author{Cheng-Tai Tan}
\email{chengtaitan@stu.pku.edu.cn}
\affiliation{School of Physics, Peking University, Beijing 100871, China}

\author{Jian Wang}
\email{j.wang@sdu.edu.cn}
\affiliation{School of Physics, Shandong University, Jinan, Shandong 250100, China}
\affiliation{Center for High Energy Physics, Peking University, Beijing 100871, China}

\author{Hong-Fei Zhang}
\email{hfzhang@mail.gufe.edu.cn}
\affiliation{College of Big Data Statistics, Guizhou University of Finance and Economics, Guiyang, 550025, China}


\begin{abstract}
Compton scattering is a fundamental process in QED with broad applications, yet its theoretical description at high energies is challenged by substantial next-to-leading order (NLO) corrections arising from double-logarithmic enhancements. To address this, we report the first calculation of the next-to-next-to-leading order (NNLO) total cross section with full electron mass dependence. Our analysis reveals that the NNLO correction, albeit still containing double logarithms, is numerically small due to a suppressing prefactor. By identifying the origin of these logarithms in a kinematic regime featuring a Glauber electron exchange, we successfully resum the leading logarithmic series to all orders, obtaining a compact result in terms of a modified Bessel function. The all-order structure reveals a suppression mechanism, which explains the rapid convergence of higher-order contributions. The combination of our NNLO calculation and all-orders resummation delivers a reliable and precise prediction, poised to serve the needs of high-precision experiments in the foreseeable future.
\end{abstract}

\maketitle
\allowdisplaybreaks

{\it Introduction}.--
Compton scattering, namely the scattering of a photon with an electron, dates back to a century ago when Compton first observed the shift in photon wavelength after interacting with an electron \cite{Compton:1923zz},
and has played an important role in the development of quantum mechanics.
Nowadays, Compton scattering has been widely applied in various areas, including medical physics \cite{Mahesh1}, X-ray/Gamma-ray astronomy \cite{1993ApJS}, and quantum phenomena \cite{Tkachev1}.
At facilities like the International Linear Collider or a future Muon Collider, Compton scattering is not only an important process that can be used to test QED predictions but also a critical tool for monitoring the beam states \cite{Phinney:2007gp}.
Frontier experiments that utilize Compton scattering as a precision means to probe the fundamental interactions require extraordinarily precise theoretical predictions.

The total cross section of Compton scattering at leading order (LO) in QED  was firstly calculated by Dirac \cite{Dirac:1926wa}, Gordon \cite{Gordon:1926emj},
and then by Klein and Nishina \cite{Klein:1929lcc} with full spin and relativistic corrections.
The next-to-leading order (NLO) QED correction has been studied both numerically and analytically \cite{Brown:1952eu,mandl1952theory,Swartz:1997im,Denner:1998nk,Lee:2020obg,Lee:2021iid}, with the high-energy behavior being analyzed in \cite{Gell-Mann:1964aya,McCoy:1976ff,Sen:1982xv,Lee:2021iid}. 
It is found that the NLO correction is substantial, ranging from 8.2\% at center-of-mass energy $\sqrt{s}=1$ GeV to around 30\% at $\sqrt{s}=1$ TeV, which largely surpasses the expectation of normal perturbative expansion in QED, i.e., $\mathcal{O}(\alpha)$ with $\alpha\sim 1/137$ being the electromagnetic coupling. 

This result can be understood from the analytical form of the total cross section at high energies \cite{Lee:2021iid},
\begin{equation}
    \sigma_{\mathrm{tot}}\sim\frac{2\pi\alpha^2}{s}\ln\left(\frac{s}{m^2}\right)\left[1+\frac{\alpha}{6\pi}\ln^2\left(\frac{s}{m^2}\right)+\cdots \right],
    \label{eq:nlo}
\end{equation}
with $m$ the electron mass. The double-logarithmic enhancement renders $\alpha \ln^2(s/m^2)$ of order $\mathcal{O}(1)$ at $\sqrt{s} \sim 1\text{ GeV}$, which eliminates the suppression of the coupling $\alpha$ at high orders. Consequently, the observed suppression at NLO stems purely from the $1/(6\pi)$ prefactor.
This large logarithmic structure prompts two immediate questions: (i) whether the next-to-next-to-leading-order (NNLO) contribution still offers a significant improvement, and 
(ii) what is the origin of these double logarithms and how to resum them to all orders?

The double logarithms are usually related to uncancelled soft-collinear overlapping singularities. 
But the total cross section of Compton scattering is free of such singularities due to the Block-Nordsieck theorem \cite{Bloch:1937pw}.
Naively thinking, it only contains a singularity in the limit when the outgoing photon becomes collinear to the ingoing electron, a process called backward scattering.
This corresponds to a Regge limit, i.e., $s\gg |t|$, which has been studied extensively \cite{Frye:2018xjj,Hannesdottir:2019opa,Hannesdottir:2019umk,DelDuca:2001gu,DelDuca:2013ara,DelDuca:2014cya,DelDuca:2011ae,Pancheri:2016yel,Lipatov:1976zz,Kuraev:1976ge,Fadin:1975cb,Balitsky:1978ic}.
The resummed leading logarithms (LLs) in the Regge limit, however, are only in the form of $(\alpha_s \ln (s/(-t))^n$,
and thus cannot be applied to the double logarithms in Compton scattering.
We note that the double logarithmic terms in QED $2\to2$ processes  at the amplitude level have been studied almost 60 years ago~\cite{Gorshkov:1966qd, Gorshkov:1966ht, Gorshkov:1966hu}, which are, however, different from the result at the cross section level. 

It is the purpose of this Letter to address the above two questions.
We calculate the NNLO total
cross section of Compton scattering by utilizing the modern multiloop techniques, which reveals further double logarithms at the high-energy limit. However, the leading logarithmic term is suppressed by a prefactor of $1/(96\pi^2)$. 
Based on the region analysis, we find that these logarithms emerge in the region with an exchange of a Glauber electron between a bunch of collinear and a bunch of anti-collinear particles.
The phase space and squared amplitude simplify significantly in this region so that we are able to predict the LLs at any order.
The all-order structure clearly explains the suppressing prefactor through double factorial terms in the denominator, indicating rapid perturbative convergence for this process.

{\it NNLO Calculation\label{sec:fix}}.--
We focus on QED NNLO corrections to the total cross section of Compton scattering 
\begin{equation}
    e^-(p_1)\gamma(p_2)\rightarrow e^-(p_3)\gamma(p_4) 
\end{equation}
with full electron mass dependence. 
We do not consider contributions from final states with three charged particles because these processes can be separated experimentally.
It is useful to define a dimensionless variable
\begin{equation}
     \tau=\frac{m^2}{s}
\end{equation}
with $s=(p_1+p_2)^2$.

We employ {\tt FeynArts} \cite{Hahn:2000kx} to generate the Feynman amplitudes and use {\tt {CalcLoop}} package \cite{calcloop} to handle Lorentz and Dirac algebras, reduce squared amplitudes to linear combinations of scalar integrals, and map the latter to predefined integral families.
To apply the modern multi-loop calculation techniques, we have treated the phase space integrals as loop integrals by virtue of reverse unitarity \cite{Anastasiou:2002yz,Anastasiou:2002qz,Anastasiou:2003yy}. 
The scalar integrals are then reduced to master integrals by using integration-by-part (IBP) identities \cite{Chetyrkin:1981qh}  based on the Laporta algorithm \cite{Laporta:2000dsw}. We use the {\tt Blade} package \cite{Guan:2024byi} to generate IBP identities and solve the IBP systems with the strategy of block-triangular form \cite{Liu:2018dmc,Guan:2019bcx}, employing {\tt FiniteFlow}~\cite{Peraro:2019svx} as a solver.
Finally, we construct the differential equations \cite{Kotikov:1990kg} of the master integrals with respect to $\tau$. 
We use the dimensional regularization with $\epsilon=(4-d)/2$ to regulate both ultraviolet (UV) and infrared divergences.
Instead of reconstructing the analytical $\epsilon$ dependence, we employ the strategy proposed in Refs. \cite{Liu:2022chg,Chen:2022vzo} to compute only a few $\epsilon$ points.  In our calculation, we take the values as $\epsilon=-1/1000$ and $\epsilon=1/1000$ \cite{Bi:2023bnq}, combination of which results in an error of $\mathcal{O}(10^{-6})$ at the cross section level.

We fix the boundary conditions of the master integrals by the auxiliary mass flow method \cite{Liu:2017jxz,Liu:2020kpc,Liu:2021wks,Liu:2022mfb} implemented in {\tt AMFlow} \cite{Liu:2022chg}. 
In principle, one can choose an arbitrary regular point in the physical region.
We choose $\tau=1/530$ close to the high-energy endpoint $\tau=0$, which is our primary concern. 
Then, all results of the master integrals in the physical region can be derived by solving the differential equations. 
Near the singular point $\tau=0$, 
we directly solve the power series expansion \cite{Caffo:2008aw,Czakon:2008zk} to extract the coefficients of all the logarithms $\ln^n\tau, n\leq5$. 
The Frobenius series solution for the differential equations of dimensionally-regularized loop integrals has the form
\begin{equation}
    f(\epsilon,\tau)=\sum_{a,b} \tau^a \ln^b(\tau) \left(\sum_{n=0}^\infty C_{abn}(\epsilon) \tau^n\right),
\end{equation}
where  $b$ is a non-negative integer and $a$ satisfies the pattern 
\begin{equation}
    a = a_0 + a_1 \epsilon,\quad 2a_0,a_1\in Z.
\end{equation}

The infrared divergences cancel after real and virtual contributions are combined together.
The UV divergence is  renormalized in the on-shell scheme. 
We adopt two methods to renormalize the bare cross sections. The first is to calculate the counterterm Feynman diagrams generated by {\tt FeynArts} and the second is to replace the bare quantities with renormalized ones. 
In the second method, the bare results are multiplied with the coupling constant renormalization $Z_\alpha^{\rm OS}$ and wave function renormalizations $Z_2^{\rm OS}$,$Z_3^{\rm OS}$ for the electron and photon, respectively.
The bare electron mass in internal propagators is replaced with the on-shell mass $m^b=m^{\rm OS}(1+\delta_m)$.
The  needed renormalization constants can be found in Refs. \cite{Broadhurst:1991fy,Melnikov:2000zc}.
After these manipulations, the cross section is expanded in the renormalized  coupling constant $\alpha$ properly.
The results of the two methods agree with each other.

At the end, the cross section in the full physical region $\tau\in (0,1)$ is obtained in terms of a piecewise power series with up to 16 segments.
In each expansion point, we deeply expand the power series to guarantee the high precision of the numerical result. We check and confirm our result at another regular point $\tau=49/53$  by direct numerical computation with {\tt AMFlow}.

The total cross section near the threshold is given by
\begin{align}
    \sigma(x) &= \frac{\pi\alpha^2}{m^2}\left(\frac{8}{3}-\frac{8}{3}x+\frac{52}{15}x^2 - \frac{133}{30}x^3 + \frac{572}{105}x^4\right) \notag\\
    &+\frac{\alpha^3}{m^2}x^2\left(-\frac{16}{9}\ln x+\frac{4}{3}x \ln x-\frac{25}{9}x^2\ln x \right.\notag\\
    & \quad  \quad  \quad \left. +\frac{7}{15} -\frac{113}{90} x  + \frac{203}{125} x^2\right)\notag\\
    &+\frac{\alpha^4}{\pi m^2}x^2(0.814x^2\ln^2 x+0.797x^2\ln x \notag\\
    &\quad  \quad  \quad-2.05+7.16x-18.6x^2)\notag\\
    &+\mathcal{O}(x^5),
\end{align}
where $x=(1-\tau)/\tau$.
The higher-order corrections are vanishing due to the quadratic dependence on $x$.
This is in concordance with  Thirring's theorem \cite{Denner:1998nk,Thirring:1950cy,Dittmaier:1997dx},
which provides a strong validation of our calculation. 
The above expression can be used to calculate the cross section when $0.9<\tau< 1$.

At the high-energy limit, i.e., $\tau\le 0.01$, we provide a deep expansion with a precision that is enough for experimental needs,
\begin{widetext}
\begin{align}
     \sigma(\tau) = &\frac{\pi\alpha^2}{s}\left[(2 L+1) + \tau(-6L+17) +\tau^2 (-30L+32) +\tau^3(-70L+48)+\tau^4(-126L+64)+\mathcal{O}(\tau^5)\right]  \notag\\
     &  +\frac{\alpha^3}{s}\left[\left(\frac{1}{3}L^3-\frac{1}{2}L^2+\frac{17}{4}L-9.5016\right) +\right. \tau(2L^3+13L^2-36.5340L+0.55139)+\notag\\  
     &\left. \hspace{20pt} +\tau^2\left(\frac{38}{3}L^3 +\frac{151}{2}L^2-89.091L+47.062\right) +\tau^3\left(\frac{157}{3}L^3+\frac{344}{3}L^2-220.07L+149.73\right)+\mathcal{O}(\tau^4)\right]  \notag\\
     & + \frac {\alpha^4} {\pi s}\bigg[ \left(\frac {1} {48} L^5 - 0.149 L^4 + 1.34 L^3 - 1.17 L^2 - 16.6 L + 23.8\right) \notag\\ 
     & \hspace{25pt}+\tau (1.13 L^5 - 6.54 L^4 - 44.6 L^3 + 442 L^2 - 36.0 L - 2.64\times 10^3) +\mathcal{O}(\tau^2)\bigg],
     \label{eq:highenergy}
\end{align}
\end{widetext}
with $L=\ln(1/\tau)$. 
The LO and NLO results agree with those given in \cite{Lee:2021iid}, while the NNLO result is new. 
We indeed find a double-logarithmic enhancement at NNLO.
Crucially, however, its leading-logarithmic coefficient is suppressed by a factor of 
$1/(96\pi^2)$, relative to the LO coefficient, significantly limiting its impact on the cross section.
Numerically, the NNLO correction improves the NLO cross section by 0.2\% at $\sqrt{s}=1$ GeV and 2.5\% at $\sqrt{s}=1$ TeV, respectively.

For $0.01< \tau\leq 0.9$, we provide a list of numerical values in the auxiliary file, 
from which the NNLO correction can be computed using interpolation.

{\it All-Order Leading Logarithms.}--
\label{sec:resum}
Now we derive the LL of Compton scattering in the high-energy limit to all orders. The logarithm at LO comes from the phase space integration of the $t$-channel diagram in the backward scattering limit.
It can also be considered as the imaginary part of the corresponding forward scattering loop amplitude.
Using the method of regions \cite{Smirnov:1997gx,Beneke:1997zp},
we find four regions: the hard ($h$), collinear ($c$), anti-collinear ($\bar{c}$), and soft ($s$) regions, as in the case of $H\to \gamma\gamma$ studied in Ref. \cite{Hou:2025ovb}.
Defining two light-cone directions $n^{\mu}_-$ and $n^{\mu}_+$ along the momenta $p_1$ and $p_2$, respectively, any momentum can be decomposed as $p^{\mu}=n_+p n^{\mu}_-/2+n_-p n^{\mu}_+/2+p^{\mu}_{\perp}=(n_+p,n_-p,p_{\perp}) $.
The four regions correspond to the ($t$-channel electron) loop momentum $q$  scaling as $\sqrt{s}(1,1,1),\sqrt{s}(1,\lambda^2,\lambda), \sqrt{s}(\lambda^2,1,\lambda) $  and $\sqrt{s}(\lambda,\lambda,\lambda)$, respectively, with the power counting parameter $\lambda=\sqrt{\tau}$.

Direct computation of the loop amplitude in the (anti-)collinear and soft regions suffers from the problem of endpoint divergences.
Imposing an analytic regulator $\nu^{\eta}/(n_+q + n_-q + i0)^{\eta}$ to regulate such divergences, we obtain the imaginary part from only the hard and anti-collinear regions.
Specifically, the hard and anti-collinear region contributions combine in the form
\begin{align}
  & \textrm{Im} \left[ \left( \frac{\mu^2}{-s-i0}\right)^{\epsilon}\frac{-1}{\epsilon^2} +  \left( \frac{\mu^2}{m^2}\right)^{\epsilon}\left( \frac{\nu}{-\sqrt{s}+i0}\right)^{\eta}\frac{1}{\epsilon \eta} \right]\nonumber\\
  =&-\pi \left[ \ln\frac{\mu^2}{s} -\ln\frac{\mu^2}{m^2}   \right] = -\pi \ln \tau\,.
  \label{eq:im}
\end{align}
This result explains the origin of the logarithmic structure clearly.
However, it seems redundant to introduce the analytic regulator since we do not need the cancellation of $1/\eta$ poles as in other cases \cite{Bell:2022ott,Hou:2025ovb}.

This regulator simply plays the role of inducing an imaginary part.
Actually, the imaginary part can be obtained more intuitively from the unitarity of the $S$-matrix.
We employ the Cutkosky rule in the anti-collinear region, forcing the photon propagator to be on-shell,
i.e., taking $\delta(n_-q\sqrt{s})$,
which cancels the $n_-q$ integration.
The $n_+q$ integration is also trivial due to the $\delta$ function associating with the on-shell electron propagator.
Then the integral collapses to the perpendicular space $q_{\perp}$ integration,
which generates the imaginary part in the second line of Eq. (\ref{eq:im}) directly.
Therefore, 
applying the Cutkosky rule turns the loop momentum $q$ from an anti-collinear mode into a Glauber mode $(G)$, which scales as $\sqrt{s}(\lambda^2,\lambda^2,\lambda)$.
In other words,
the Glauber region is a subset of the anti-collinear region.
It is only this region that could yield an imaginary part \cite{Donoghue:2014mpa}.

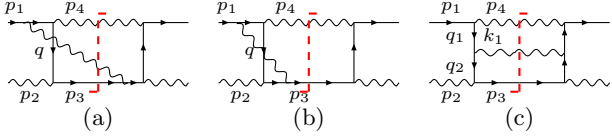
\begin{figure}[htbp]
\centering

\begin{tikzpicture}[scale=0.4]

    \begin{scope}[shift={(7,0)}]
    \begin{feynman}

        \vertex (a1) at (-3, 1);
        \vertex (a2) at (-3, -1);
        \vertex (b1) at (-1.5, 1);
        \vertex (b2) at (-1.5, -1);
        \vertex (c1) at (-0., 1);
        \vertex (c2) at (-0., -1);
        
        \vertex (d1) at (0., 1);
        \vertex (d2) at (0., -1);
        \vertex (e1) at (1.5, 1);
        \vertex (e2) at (1.5, -1);
        \vertex (f1) at (3, 1);
        \vertex (f2) at (3, -1);
        
        \vertex (left_mid) at (-1.5, 0);
        \vertex (right_mid) at (1.5, 0);
        \vertex (mid) at (-0.2,0);
        
        \diagram* {
            (a1) -- [fermion, edge label={\scriptsize \(p_1\)}, arrow size=0.2mm] (b1),
            (a2) -- [photon, edge label'={\scriptsize \(p_2\)}] (b2),
            (b1) -- [fermion, edge label'={\scriptsize \(q_1\)}, arrow size=0.2mm] (left_mid),
            (left_mid) -- [fermion, edge label'={\scriptsize \(q_2\)}, arrow size=0.2mm] (b2),
            (b1) -- [photon, edge label={\scriptsize \(p_4\)}, arrow size=0.2mm] (c1),
            (b2) -- [fermion, edge label'={\scriptsize \(p_3\)}, arrow size=0.2mm] (c2),
        };
        
        \diagram* {
            (d1) -- [photon] (e1),
            (d2) -- [fermion, arrow size=0.2mm] (e2),
            (e1) -- [anti fermion, arrow size=0.2mm] (right_mid),
            (right_mid) -- [anti fermion, arrow size=0.2mm] (e2),
            (e1) -- [fermion, arrow size=0.2mm] (f1),
            (e2) -- [photon] (f2),
        };
        
        \diagram* {
            (left_mid) -- [photon, edge label={\scriptsize \(k_1\)}] (mid),
            (mid) -- [photon] (right_mid),
        };
        
        \draw[ thick, red, dashed] (0, -1.3) -- (0, 1.3);
        \draw[ thick, red, dashed] (0, 1.3) -- (0.3, 1.3);
        \draw[ thick, red, dashed] (-0.3, -1.3) -- (0, -1.3);
        
    \end{feynman}
    \end{scope}
    
    \begin{scope}[shift={(-7,0)}]
    \begin{feynman}

        \vertex (a1) at (-3, 1);
        \vertex (a2) at (-3, -1);
        \vertex (b1) at (-1.5, 1);
        \vertex (b2) at (-1.5, -1);
        \vertex (c1) at (-0., 1);
        \vertex (c2) at (-0., -1);

        \vertex (d1) at (0., 1);
        \vertex (d2) at (0., -1);
        \vertex (e1) at (1.5, 1);
        \vertex (e2) at (1.5, -1);
        \vertex (f1) at (3, 1);
        \vertex (f2) at (3, -1);

        \vertex (left_ein_mid) at (-2.5, 1);

        \vertex (right_ein_mid) at (1, -1);
        
        \diagram* {
            (a1) -- [fermion, edge label={\scriptsize \(p_1\)}, arrow size=0.2mm] (left_ein_mid),
            (left_ein_mid) -- [fermion, arrow size=0.2mm] (b1),
            (a2) -- [photon, edge label'={\scriptsize \(p_2\)}] (b2),
            (b1) -- [fermion, arrow size=0.2mm, edge label'={\scriptsize \(q\)}] (b2),
            (b1) -- [photon, edge label={\scriptsize \(p_4\)}] (c1),
            (b2) -- [fermion, edge label'={\scriptsize \(p_3\)}, arrow size=0.2mm] (c2),
        };

        \diagram* {
            (d1) -- [photon] (e1),
            (d2) -- [fermion, arrow size=0.2mm] (right_ein_mid),
            (right_ein_mid) -- [fermion, arrow size=0.2mm] (e2),
            (e1) -- [anti fermion, arrow size=0.2mm] (e2),
            (e1) -- [fermion, arrow size=0.2mm] (f1),
            (e2) -- [photon, arrow size=0.2mm] (f2),
        };

        \diagram* {
            (left_ein_mid) -- [photon] (right_ein_mid),
        };

        \draw[thick, red, dashed] (0, -1.3) -- (0, 1.3);
        \draw[thick, red, dashed] (0, 1.3) -- (0.3, 1.3);
        \draw[thick, red, dashed] (-0.3, -1.3) -- (0, -1.3);

    \end{feynman}
    \end{scope}

    \begin{scope}[shift={(0,0)}]
    \begin{feynman}

        \vertex (a1) at (-3, 1);
        \vertex (a2) at (-3, -1);
        \vertex (b1) at (-1.5, 1);
        \vertex (b2) at (-1.5, -1);
        \vertex (c1) at (-0., 1);
        \vertex (c2) at (-0., -1);

        \vertex (d1) at (0., 1);
        \vertex (d2) at (0., -1);
        \vertex (e1) at (1.5, 1);
        \vertex (e2) at (1.5, -1);
        \vertex (f1) at (3, 1);
        \vertex (f2) at (3, -1);

        \vertex (ein_mid) at (-2.4, 1);

        \vertex (eout_mid) at (-0.6, -1);

        \diagram* {
            (a1) -- [fermion, edge label={\scriptsize \(p_1\)}, arrow size=0.2mm] (ein_mid),
            (ein_mid) -- [fermion, arrow size=0.2mm] (b1),
            (a2) -- [photon, edge label'={\scriptsize \(p_2\)}] (b2),
            (b1) -- [fermion, edge label'={\scriptsize \(q\)}, arrow size=0.2mm] (b2),
            (b1) -- [photon, edge label={\scriptsize \(p_4\)}] (c1),
            (b2) -- [fermion, arrow size=0.2mm] (eout_mid),
            (eout_mid) -- [fermion, edge label'={\scriptsize \(p_3\)}, arrow size=0.2mm] (c2),
        };

        \diagram* {
            (d1) -- [photon] (e1),
            (d2) -- [fermion, arrow size=0.2mm] (e2),
            (e1) -- [anti fermion, arrow size=0.2mm] (e2),
            (e1) -- [fermion, arrow size=0.2mm] (f1),
            (e2) -- [photon] (f2),
        };

        \diagram* {
            (ein_mid) -- [photon] (eout_mid),
        };

        \draw[ thick, red, dashed] (0, -1.3) -- (0, 1.3);
        \draw[ thick, red, dashed] (0, 1.3) -- (0.3, 1.3);
        \draw[ thick, red, dashed] (-0.3, -1.3) -- (0, -1.3);

    \end{feynman}
    \end{scope}

    \node at (-7, -2.2) {(a)};
    \node at (0, -2.2) {(b)};
    \node at (7, -2.2) {(c)};
\end{tikzpicture}

\caption{NLO Feynman diagrams contributing to LLs. }
\label{fig:nlo}
\end{figure}

At NLO, the extra double logarithm can usually be caused by the soft and collinear singularities induced by the additional real or virtual photon. 
If the additional photon is radiated from the external electrons, the double logarithms always cancel between the  corresponding virtual and real corrections which share the same topology; see an example in Fig. \ref{fig:nlo}(a) and (b). 
Consequently, only the diagram with the additional photon emitted from the $t$-channel electron propagator, as shown in Fig. \ref{fig:nlo}(c), could contribute to double logarithms.
The full result of this diagram in the Feynman gauge is given by
\begin{align}
\sigma_{\mathrm{ladder}}^{\mathrm{NLO}}=& \frac{\alpha^3}{s}\bigg(\frac{L}{\epsilon}+\frac{1}{3}L^3 +\frac{1}{2}L^2+L+ \notag\\
&\frac{5}{2}-\frac{2\pi^2}{3}-2\zeta_3+\mathcal{O}(\tau)\bigg).
\end{align}
The above IR divergence $1/\epsilon$ would cancel after including the renormalization of the electron wave function, which also contains IR divergences in the on-shell scheme.

The above LL is the same as that in the NLO total cross section in Eq. (\ref{eq:highenergy}).
The relevant regions for the imaginary part of the forward scattering amplitude consist of $(q_1,q_2)\sim (h,h),(c,h),(h,\bar{c}), (c,G), (G,\bar{c}) $ with $q_i$ the momenta of electron propagators; see Fig. \ref{fig:nlo}(c).
Note that the overlap of $(c,G)$ and $ (G,\bar{c}) $ regions is nonvanishing and thus has to be subtracted.
After finding all the contributing regions, one can set the scale $\mu=\sqrt{s}$ so that the regions containing $h$ do not produce LLs.
All the LLs come only from $(c,G)$ or $ (G,\bar{c}) $
or their overlap.
This judicious choice of scale would simplify the calculation a lot.

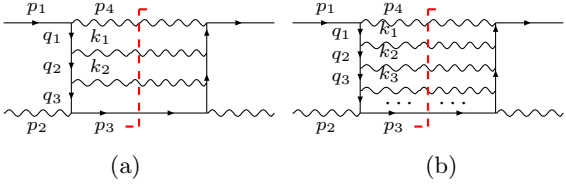
\begin{figure}[htbp]
\centering

\begin{tikzpicture}[scale=0.6]

    \begin{scope}[shift={(-3.2,0)}]
    \begin{feynman}

        \vertex (a1) at (-3, 1);
        \vertex (a2) at (-3, -1);
        \vertex (b1) at (-1.5, 1);
        \vertex (b2) at (-1.5, -1);
        \vertex (c1) at (-0., 1);
        \vertex (c2) at (-0., -1);
        
        \vertex (d1) at (0., 1);
        \vertex (d2) at (0., -1);
        \vertex (e1) at (1.5, 1);
        \vertex (e2) at (1.5, -1);
        \vertex (f1) at (3, 1);
        \vertex (f2) at (3, -1);
        
        \vertex (left_mid) at (-1.5, 0.33);
        \vertex (right_mid) at (1.5, 0.33);
        \vertex (mid) at (-0.2,0.33);

        \vertex (left_mid2) at (-1.5, -0.33);
        \vertex (right_mid2) at (1.5, -0.33);
        \vertex (mid2) at (-0.2,-0.33);
        
        \diagram* {
            (a1) -- [fermion, edge label={\scriptsize \(p_1\)}, arrow size=0.2mm] (b1),
            (a2) -- [photon, edge label'={\scriptsize \(p_2\)}] (b2),
            (b1) -- [fermion, edge label'={\scriptsize \(q_1\)}, arrow size=0.2mm] (left_mid),
            (left_mid) -- [fermion, edge label'={\scriptsize \(q_2\)}, arrow size=0.2mm] (left_mid2),
            (left_mid2) -- [fermion, edge label'={\scriptsize \(q_3\)}, arrow size=0.2mm] (b2),
            (b1) -- [photon, edge label={\scriptsize \(p_4\)}, arrow size=0.2mm] (c1),
            (b2) -- [fermion, edge label'={\scriptsize \(p_3\)}, arrow size=0.2mm] (c2),
        };
        
        \diagram* {
            (d1) -- [photon] (e1),
            (d2) -- [fermion, arrow size=0.2mm] (e2),
            (e1) -- [anti fermion, arrow size=0.2mm] (right_mid),
            (right_mid) -- [anti fermion, arrow size=0.2mm] (e2),
            (e1) -- [fermion, arrow size=0.2mm] (f1),
            (e2) -- [photon] (f2),
        };
        
        \diagram* {
            (left_mid) -- [photon, edge label={\scriptsize \(k_1\)}] (mid),
            (mid) -- [photon] (right_mid),
        };

        \diagram* {
            (left_mid2) -- [photon, edge label={\scriptsize \(k_2\)}] (mid2),
            (mid2) -- [photon] (right_mid2),
        };
        
        \draw[ thick, red, dashed] (0, -1.3) -- (0, 1.3);
        \draw[ thick, red, dashed] (0, 1.3) -- (0.3, 1.3);
        \draw[ thick, red, dashed] (-0.3, -1.3) -- (0, -1.3);
        
    \end{feynman}
    \end{scope}

    \begin{scope}[shift={(3.2,0)}]
    \begin{feynman}

        \vertex (a1) at (-3, 1);
        \vertex (a2) at (-3, -1);
        \vertex (b1) at (-1.5, 1);
        \vertex (b2) at (-1.5, -1);
        \vertex (c1) at (-0., 1);
        \vertex (c2) at (-0., -1);
        
        \vertex (d1) at (0., 1);
        \vertex (d2) at (0., -1);
        \vertex (e1) at (1.5, 1);
        \vertex (e2) at (1.5, -1);
        \vertex (f1) at (3, 1);
        \vertex (f2) at (3, -1);
        
        \vertex (left_mid) at (-1.5, 0.5);
        \vertex (right_mid) at (1.5, 0.5);
        \vertex (mid) at (-0.2,0.5);

        \vertex (left_mid2) at (-1.5, 0);
        \vertex (right_mid2) at (1.5, 0);
        \vertex (mid2) at (-0.2,0);

        \vertex (left_mid3) at (-1.5, -0.5);
        \vertex (right_mid3) at (1.5, -0.5);
        \vertex (mid3) at (-0.2,-0.5);
        
        \diagram* {
            (a1) -- [fermion, edge label={\scriptsize \(p_1\)}, arrow size=0.2mm] (b1),
            (a2) -- [photon, edge label'={\scriptsize \(p_2\)}] (b2),
            (b1) -- [fermion, edge label'={\scriptsize \(q_1\)}, arrow size=0.2mm] (left_mid),
            (left_mid) -- [fermion, edge label'={\scriptsize \(q_2\)}, arrow size=0.2mm] (left_mid2),
            (left_mid2) -- [fermion, edge label'={\scriptsize \(q_3\)}, arrow size=0.2mm] (left_mid3),
            (left_mid3) -- [fermion, arrow size=0.2mm] (b2),
            (b1) -- [photon, edge label={\scriptsize \(p_4\)}, arrow size=0.2mm] (c1),
            (b2) -- [fermion, edge label'={\scriptsize \(p_3\)}, arrow size=0.2mm] (c2),
        };
        
        \diagram* {
            (d1) -- [photon] (e1),
            (d2) -- [fermion, arrow size=0.2mm] (e2),
            (e1) -- [anti fermion, arrow size=0.2mm] (right_mid),
            (right_mid) -- [anti fermion, arrow size=0.2mm] (e2),
            (e1) -- [fermion, arrow size=0.2mm] (f1),
            (e2) -- [photon] (f2),
        };
        
        \diagram* {
            (left_mid) -- [photon, edge label={\scriptsize \(k_1\)}] (mid),
            (mid) -- [photon] (right_mid),
        };

        \diagram* {
            (left_mid2) -- [photon, edge label={\scriptsize \(k_2\)}] (mid2),
            (mid2) -- [photon] (right_mid2),
        };

        \diagram* {
            (left_mid3) -- [photon, edge label={\scriptsize \(k_3\)}] (mid3),
            (mid3) -- [photon] (right_mid3),
        };
        
        \draw[ thick, red, dashed] (0, -1.3) -- (0, 1.3);
        \draw[ thick, red, dashed] (0, 1.3) -- (0.3, 1.3);
        \draw[ thick, red, dashed] (-0.3, -1.3) -- (0, -1.3);

        \node at ( -0.6, -0.8) {$\cdots$};
        \node at ( 0.6, -0.8) {$\cdots$};
        
     \end{feynman}
     \end{scope}
   
    \node at (-3.5, -2.2) {(a)};
    \node at (3.5, -2.2) {(b)};
\end{tikzpicture}

\caption{NNLO and all-order ladder diagrams.}
\label{fig:nnlo}
\end{figure}

These structures can be extended to all orders.
The $t$-channel electron momenta of the $n$-loop ladder diagram in the N$^{n-1}$LO correction are labeled as $q_1,q_2,\cdots,q_n$ respectively, as shown in Fig. \ref{fig:nnlo}.
We adopt the notation $r_i=(c,\cdots,c,G,\bar{c},\cdots,\bar{c})$ to denote the region where $q_j \,(j<i)$ and $q_k \,(k>i)$ are collinear and anti-collinear, respectively, while $q_i$ is a Glauber mode.
The momenta of final-state photons are denoted by $k_i, i=0,\cdots,n-1,$ with $k_0=p_4$. 
Momentum conservation requires $k_i = q_i-q_{i+1}$ with $q_0=p_1$,
and $p_3=q_{n}+p_2$.
The LL arises when the light-cone momentum components obey the strong ordering
\begin{subequations}
\begin{align}
    n_+q_0 > n_+q_1 >\cdots>n_+q_n>0\,,
  \label{eq:ordering1} \\
   0<-n_-q_1 <\cdots < -n_-q_n<n_-p_2  \,.
  \label{eq:ordering2}
\end{align}  
  \label{eq:ordering}
\end{subequations}
The first equation (\ref{eq:ordering1}) always holds since $n_{\pm}k_i>0$ and $k_i^2=0$, while the second equation (\ref{eq:ordering2}) helps chisel out the subregion contributing to LLs.
It is easy to check that this subregion is contained in any of the $r_i$ regions.
Therefore, all $r_i$ regions as well as any overlap give the same result.

The $(n+1)$-body phase space can be written as 
\begin{align}
    &\int\mathrm{dPS}_{n+1} = (2\pi)\int \prod_{i=1}^n\frac{\mathrm{d}^4 q_i}{(2\pi)^{3}}\delta^+(p_3^2-m^2)\prod_{j=0}^{n-1}\delta^+(k_j^2)\,
    \notag
\end{align}
with 
$\delta^+(p^2)=\delta(p^2)\theta(p^0).$
Based on the strong ordering in (\ref{eq:ordering}) and a useful trick in  dealing with LLs, i.e., dropping the smaller component in an expression,
the on-shell conditions are given by $\delta(k_0^2)\approx \delta((n_+q_0)( n_-q_0 -n_-q_{1})-q_{1T}^2)$, $\delta(k_i^2)\approx \delta(-(n_+q_i)( n_-q_{i+1})-(q_i-q_{i+1})_T^2),i=1,\cdots, n-1,$ and $\delta(p_3^2-m^2)\approx \delta(n_+q_n n_-p_2-q_{nT}^2-m^2)$,
which render the integrations of $n_-q_i$ ($i=1,\cdots,n$) and $n_+q_n$ trivial.
Simultaneously, the squared amplitude is also simplified
with the denominators $q_i^2-m^2 \approx -q_{iT}^2-m^2-\frac{n_+q_i}{n_+q_{i-1}}q_{(i-1)T}^2.$
Defining the variables $x_i=n_+q_i \sqrt{s}/m^2$ and $y_i = (q_{iT}^2+m^2)/m^2$, the LL in the cross section is given by
\begin{align}
    &\sigma^{\mathrm{ N}^{n-1}{\rm LO}}_{\mathrm {LL}}=\frac{4\pi^2}{s} \frac{\alpha^{n+1}}{(2\pi)^{n}}
    \int_{1}^{1/\tau}\frac{\mathrm{d}x_1}{x_1}
    \prod_{j=2}^{n-1}\int_{1}^{x_{j-1}}\frac{\mathrm{d}x_j}{x_j}\notag\\
    &\times \int_{1}^{x_1}\frac{\mathrm{d}y_1}{y_1}\prod_{i=2}^{n-1}\int_{\mathrm{max}\{1,\frac{x_i}{x_{i-1}}y_{i-1}\}}^{x_i}\frac{\mathrm{d}y_i}{y_i} \int_{1}^{x_{n-1}}\frac{\mathrm{d}y_{n}}{y_{n}},
    \label{eq:iteint}
\end{align}
where we have replaced the dimensional regularization of the divergences by a cutoff regularization.
The lower bound of $x_j$ and upper bound of $y_j$ with  $j=1,\cdots,n-1$ are determined since $q_j$ is supposed to be a collinear momentum.  
The upper bound of $y_n$ is set because of $y_n=x_n<x_{n-1}$.
The other bounds can be understood easily.
These $2n-1$ fold integrations generate the $(2n-1)$-th  powers of logarithms.
The non-ladder diagrams do not have this recursive structure and thus contain no LLs.

Evaluating the above integral leads to the result
\begin{equation}
    \sigma^{\mathrm{ N}^{n-1}{\rm LO}}_{\mathrm {LL}} = -\frac{2\alpha^{n+1}}{(2\pi)^{n-2}s}\frac{1}{(n-1)!(n+1)!}\ln^{2n-1}\tau.
    \label{eq:LL}
\end{equation}
This formula agrees with the NLO and NNLO results in Eq.~\eqref{eq:highenergy}.
To further validate the result, we calculated the NNNLO ladder diagram using the multiloop techniques described above, and found full agreement. The double factorials in the denominator lead to a strong numerical suppression at high orders, explaining why NNLO contributions are not so prominent for Compton scattering in the high-energy limit where large logarithmic enhancement exists.

Summing the all order corrections, we obtain the LLs resummed cross section
\begin{align}
    \sigma_{\rm LL} &= -\frac{8\pi^2 \alpha}{s\ln \tau}I_2\left( \sqrt{\frac{2\alpha}{\pi}}\ln\tau\right)\nonumber\\
    &=\frac{2^{9/4}\pi^{7/4} \alpha^{3/4}}{ s( -\ln \tau )^{3/2} ~\tau^{\sqrt{2\alpha/\pi} } }  \quad \textrm{as} \quad \tau\to 0,
\end{align}
with $I_2(z)$ being the modified Bessel function of the first kind.
In the second line of the above equation, we present the asymptotic expansion of the resummed result in the limit of $\tau\to 0$.
For fixed $m$, the cross section is scaling as $s^{-1+\sqrt{2\alpha/\pi}}/\ln^{3/2} (s/m^2)$ as $s\to \infty$, a pattern in contrast to the Regge trajectory.
On the other hand, the cross section becomes divergent as $m\to 0$ for fixed $s$ since new infrared divergences would emerge in this limit.

\begin{figure}[htbp]
    \centering
    \hspace{-30pt}
    \includegraphics[width=0.5\textwidth]{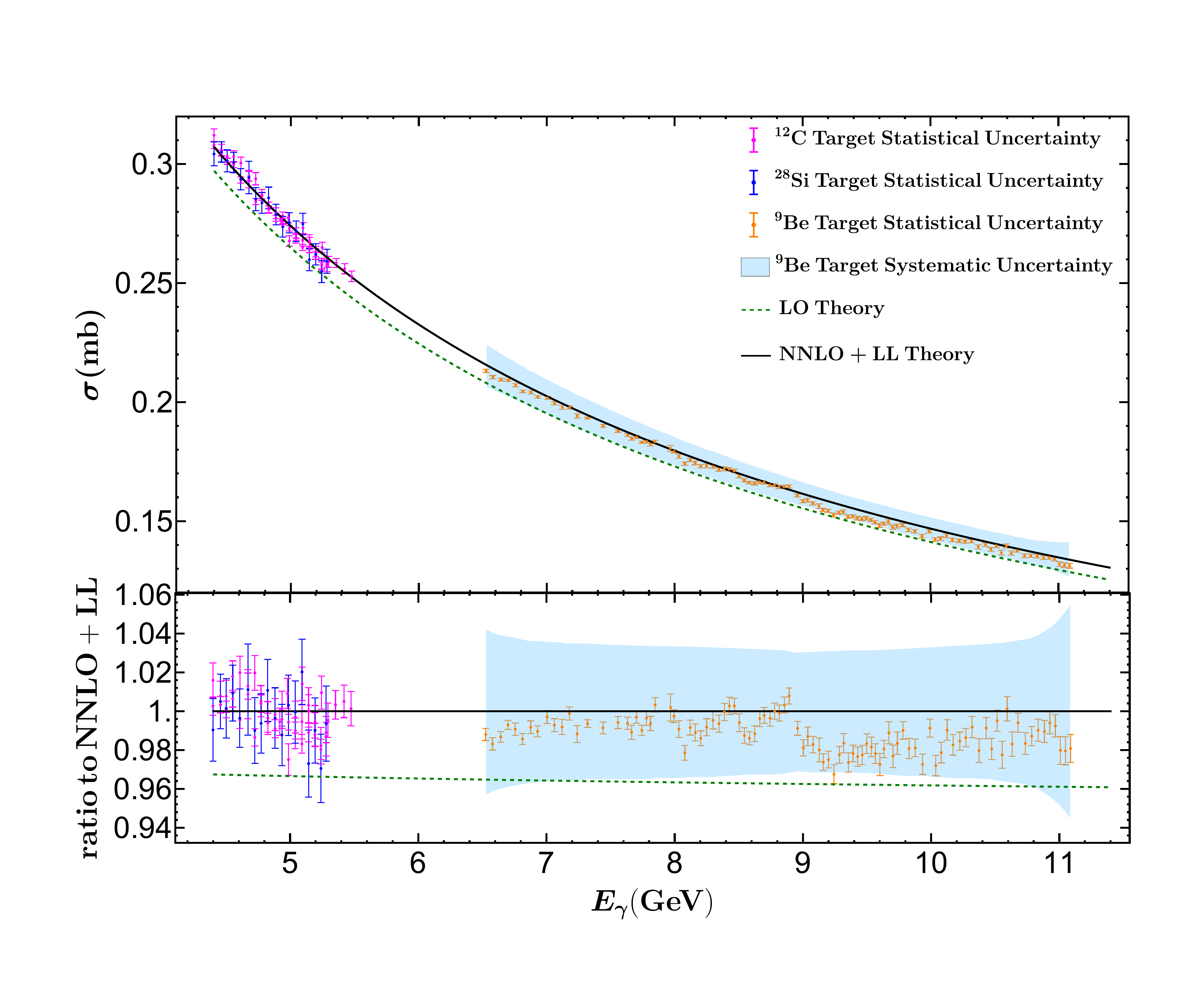}
    \vspace{-20pt}
    \caption{Comparison of theory and experiment for Compton scattering. PrimEx \cite{PrimEx:2019zre} (4.400–5.475 GeV) and GlueX \cite{GlueX:2025hve} (6.5–11.1 GeV) data for $\rm{^{12}C}$ (magenta), $\rm{^{28}Si}$ (blue), and $\rm{^{9}Be}$ (orange) are compared to theoretical predictions at LO (green dashed) and NNLO+LL (black solid). The NLO result is not shown as it is visually identical to NNLO+LL.
    }
    \label{fig:xsec}
\end{figure}

{\it Comparison with data.\label{sec:conclusion}}--
The strong suppression of LLs at higher orders ensures that our combined NNLO + resummation result provides an exceptionally precise prediction for the Compton scattering cross section. This is demonstrated phenomenologically in Fig. \ref{fig:xsec}, where we compare our most precise theoretical prediction with high-precision data from the PrimEx \cite{PrimEx:2019zre} and GlueX \cite{GlueX:2025hve} collaborations using photons with energies of 4.400–5.475 GeV and 6.5–11.1 GeV, respectively. The numerical values of the fine-structure constant and electron mass are taken from precise measurements of \cite{doi:10.1126/science.aap7706} and \cite{sturm2014high}, i.e., $\alpha = 1/137.035999$ and $m_e = 0.509108~\mathrm{MeV}$. The theory and data show excellent agreement. We emphasize that the combined NNLO and resummation effects are minuscule (at the~0.05\% level) in this energy regime, underscoring the rapid convergence of the perturbative QED series and the definitive nature of our prediction.

{\it Summary.\label{sec:conclusion}}--
We have performed a precision study of  the Compton scattering total cross section, resolving the long-standing puzzle of the size of high-order corrections, particularly the role of high-energy double logarithms. Our calculation, which provides the first complete NNLO result with full electron mass dependence, reveals that the anticipated double-logarithmic enhancement is indeed present but numerically suppressed due to a small coefficient, leading to excellent perturbative convergence.

We traced the origin of these logarithms to a specific kinematic configuration induced by a Glauber electron exchange. This insight allowed us to resum the leading logarithmic contributions to all orders in perturbation theory, yielding a compact result expressed in terms of a modified Bessel function. The all-order analysis uncovers a suppression mechanism: double factorial terms in the denominator systematically dampen the leading logarithms. The combination of our  NNLO calculation and all-orders resummation delivers a precise QED prediction, which we have validated against recent high-precision data from the PrimEx and GlueX collaborations.

The methodology developed here—particularly the identification of Glauber-type exchanges as the source of leading logarithms—opens a new avenue for high-precision calculations in other fundamental processes. This approach is directly applicable to processes like $e^+e^- \to \gamma\gamma$ and $gg \to t\bar{t}$ in the high-energy limit, where similar logarithmic structures arise from massive fermion exchanges. Our work thus provides a blueprint for achieving percent-level precision in a range of scattering processes at future colliders.

{\it Acknowledgments.}--
We would like to thank Jia-Yang Dong,  Kirill Kudashkin, and Da-Jiang Zhang for the useful discussion and collaboration on similar projects. The work was supported in part by the National Natural Science Foundation of 	China (No.12325503, No.12321005, No.12275156, No.12005117) and the High-performance Computing Platform of Peking University.
The diagrams are drawn using {\tt TikZ-Feynman} \cite{Ellis:2016jkw}.


\bibliographystyle{utphysMa}
\bibliography{ref}

\providecommand{\href}[2]{#2}\begingroup\raggedright\begin{thebibliography}{10}

\bibitem{Compton:1923zz}
A.~H. Compton, {\it {A Quantum Theory of the Scattering of X-rays by Light
  Elements}},  \href{http://dx.doi.org/10.1103/PhysRev.21.483}{{\em Phys. Rev.}
  {\bfseries 21} (1923) 483--502}
  [\href{http://inspirehep.net/search?p=find+Compton:1923zz}{{\ttfamily
  InSPIRE}}].

\bibitem{Mahesh1}
M.~Mahesh, {\it {The Essential Physics of Medical Imaging, Third Edition}},
  \href{http://dx.doi.org/10.1118/1.4811156}{{\em Med. Phys.} {\bfseries 40}
  (2013) 077301} [\href{http://inspirehep.net/search?p=find+Mahesh1}{{\ttfamily
  InSPIRE}}].

\bibitem{1993ApJS}
V.~{Schoenfelder}, H.~{Aarts}, K.~{Bennett}, H.~{de Boer}, J.~{Clear},
  W.~{Collmar}, A.~{Connors}, A.~{Deerenberg}, R.~{Diehl}, A.~{von Dordrecht},
  J.~W. {den Herder}, W.~{Hermsen}, M.~{Kippen}, L.~{Kuiper}, G.~{Lichti},
  J.~{Lockwood}, J.~{Macri}, M.~{McConnell}, D.~{Morris}, R.~{Much}, J.~{Ryan},
  G.~{Simpson}, M.~{Snelling}, G.~{Stacy}, H.~{Steinle}, A.~{Strong}, B.~N.
  {Swanenburg}, B.~{Taylor}, C.~{de Vries}, and C.~{Winkler}, {\it {Instrument
  Description and Performance of the Imaging Gamma-Ray Telescope COMPTEL aboard
  the Compton Gamma-Ray Observatory}},
  \href{http://dx.doi.org/10.1086/191794}{{\em The Astrophysical Journal
  Supplement Series} {\bfseries 86} (1993) 657}
  [\href{http://inspirehep.net/search?p=find+1993ApJS}{{\ttfamily InSPIRE}}].

\bibitem{Tkachev1}
I.~Tkachev, S.~Musin, D.~Abdurashitov, A.~Baranov, F.~Guber, A.~Ivashkin, and
  A.~Strizhak, {\it Measuring the evolution of entanglement in compton
  scattering},  \href{http://dx.doi.org/10.1038/s41598-025-87095-4}{{\em
  Scientific Reports} {\bfseries 15} (2025) 6064}
  [\href{http://inspirehep.net/search?p=find+Tkachev1}{{\ttfamily InSPIRE}}].
  \url{https://doi.org/10.1038/s41598-025-87095-4}.

\bibitem{Phinney:2007gp}
G.~Aarons {\em et al.}, {\it {ILC Reference Design Report Volume 3 -
  Accelerator}},  [\href{http://arxiv.org/abs/0712.2361}{{\ttfamily
  arXiv:0712.2361}}]
  [\href{http://inspirehep.net/search?p=find+Phinney:2007gp}{{\ttfamily
  InSPIRE}}].

\bibitem{Dirac:1926wa}
P.~A.~M. Dirac, {\it {Relativity quantum mechanics with an application to
  Compton scattering}},  \href{http://dx.doi.org/10.1098/rspa.1926.0074}{{\em
  Proc. Roy. Soc. Lond. A} {\bfseries 111} (1926) 405--423}
  [\href{http://inspirehep.net/search?p=find+Dirac:1926wa}{{\ttfamily
  InSPIRE}}].

\bibitem{Gordon:1926emj}
W.~Gordon, {\it {Der Comptoneffekt nach der Schr\"odingerschen Theorie}},
  \href{http://dx.doi.org/10.1007/BF01390840}{{\em Z. Phys.} {\bfseries 40}
  (1926) 117--133}
  [\href{http://inspirehep.net/search?p=find+Gordon:1926emj}{{\ttfamily
  InSPIRE}}].

\bibitem{Klein:1929lcc}
O.~Klein and Y.~Nishina, {\it {\"Uber die Streuung von Strahlung durch freie
  Elektronen nach der neuen relativistischen Quantendynamik von Dirac}},
  \href{http://dx.doi.org/10.1007/BF01366453}{{\em Z. Phys.} {\bfseries 52}
  (1929) 853--868}
  [\href{http://inspirehep.net/search?p=find+Klein:1929lcc}{{\ttfamily
  InSPIRE}}].

\bibitem{Brown:1952eu}
L.~M. Brown and R.~P. Feynman, {\it {Radiative corrections to Compton
  scattering}},  \href{http://dx.doi.org/10.1103/PhysRev.85.231}{{\em Phys.
  Rev.} {\bfseries 85} (1952) 231--244}
  [\href{http://inspirehep.net/search?p=find+Brown:1952eu}{{\ttfamily
  InSPIRE}}].

\bibitem{mandl1952theory}
F.~Mandl and T.~Skyrme, {\it The theory of the double compton effect},  {\em
  Proceedings of the Royal Society of London. Series A. Mathematical and
  Physical Sciences} {\bfseries 215} (1952) 497--507
  [\href{http://inspirehep.net/search?p=find+mandl1952theory}{{\ttfamily
  InSPIRE}}].

\bibitem{Swartz:1997im}
M.~L. Swartz, {\it {A Complete order alpha**3 calculation of the cross-section
  for polarized Compton scattering}},
  \href{http://dx.doi.org/10.1103/PhysRevD.58.014010}{{\em Phys. Rev. D}
  {\bfseries 58} (1998) 014010}
  [\href{http://arxiv.org/abs/hep-ph/9711447}{{\ttfamily hep-ph/9711447}}]
  [\href{http://inspirehep.net/search?p=find+Swartz:1997im}{{\ttfamily
  InSPIRE}}].

\bibitem{Denner:1998nk}
A.~Denner and S.~Dittmaier, {\it {Complete O(alpha) QED corrections to
  polarized Compton scattering}},
  \href{http://dx.doi.org/10.1016/S0550-3213(98)00767-6}{{\em Nucl. Phys. B}
  {\bfseries 540} (1999) 58--86}
  [\href{http://arxiv.org/abs/hep-ph/9805443}{{\ttfamily hep-ph/9805443}}]
  [\href{http://inspirehep.net/search?p=find+Denner:1998nk}{{\ttfamily
  InSPIRE}}].

\bibitem{Lee:2020obg}
R.~N. Lee, A.~A. Lyubyakin, and V.~A. Stotsky, {\it {Total cross sections of
  $e\gamma\to e X\bar{X}$ processes with $X=\mu,\gamma, e$ via multiloop
  methods}},  \href{http://dx.doi.org/10.1007/JHEP01(2021)144}{{\em JHEP}
  {\bfseries 01} (2021) 144} [\href{http://arxiv.org/abs/2010.15430}{{\ttfamily
  arXiv:2010.15430}}]
  [\href{http://inspirehep.net/search?p=find+Lee:2020obg}{{\ttfamily
  InSPIRE}}].

\bibitem{Lee:2021iid}
R.~N. Lee, M.~D. Schwartz, and X.~Zhang, {\it {Compton Scattering Total Cross
  Section at Next-to-Leading Order}},
  \href{http://dx.doi.org/10.1103/PhysRevLett.126.211801}{{\em Phys. Rev.
  Lett.} {\bfseries 126} (2021) 211801}
  [\href{http://arxiv.org/abs/2102.06718}{{\ttfamily arXiv:2102.06718}}]
  [\href{http://inspirehep.net/search?p=find+Lee:2021iid}{{\ttfamily
  InSPIRE}}].

\bibitem{Gell-Mann:1964aya}
M.~Gell-Mann, M.~Goldberger, F.~Low, E.~Marx, and F.~Zachariasen, {\it
  {Elementary Particles of Conventional Field Theory as Regge Poles. III}},
  \href{http://dx.doi.org/10.1103/PhysRev.133.B145}{{\em Phys. Rev.} {\bfseries
  133} (1964) B145--B160}
  [\href{http://inspirehep.net/search?p=find+Gell-Mann:1964aya}{{\ttfamily
  InSPIRE}}].

\bibitem{McCoy:1976ff}
B.~M. McCoy and T.~T. Wu, {\it {Theory of Fermion Exchange in Massive Quantum
  Electrodynamics at High-Energy. 1.}},
  \href{http://dx.doi.org/10.1103/PhysRevD.13.369}{{\em Phys. Rev. D}
  {\bfseries 13} (1976) 369--378}
  [\href{http://inspirehep.net/search?p=find+McCoy:1976ff}{{\ttfamily
  InSPIRE}}].

\bibitem{Sen:1982xv}
A.~Sen, {\it {Asymptotic Behavior of the Fermion and Gluon Exchange Amplitudes
  in Massive Quantum Electrodynamics in the Regge Limit}},
  \href{http://dx.doi.org/10.1103/PhysRevD.27.2997}{{\em Phys. Rev. D}
  {\bfseries 27} (1983) 2997}
  [\href{http://inspirehep.net/search?p=find+Sen:1982xv}{{\ttfamily InSPIRE}}].

\bibitem{Bloch:1937pw}
F.~Bloch and A.~Nordsieck, {\it {Note on the Radiation Field of the electron}},
   \href{http://dx.doi.org/10.1103/PhysRev.52.54}{{\em Phys. Rev.} {\bfseries
  52} (1937) 54--59}
  [\href{http://inspirehep.net/search?p=find+Bloch:1937pw}{{\ttfamily
  InSPIRE}}].

\bibitem{Frye:2018xjj}
C.~Frye, H.~Hannesdottir, N.~Paul, M.~D. Schwartz, and K.~Yan, {\it {Infrared
  Finiteness and Forward Scattering}},
  \href{http://dx.doi.org/10.1103/PhysRevD.99.056015}{{\em Phys. Rev. D}
  {\bfseries 99} (2019) 056015}
  [\href{http://arxiv.org/abs/1810.10022}{{\ttfamily arXiv:1810.10022}}]
  [\href{http://inspirehep.net/search?p=find+Frye:2018xjj}{{\ttfamily
  InSPIRE}}].

\bibitem{Hannesdottir:2019opa}
H.~Hannesdottir and M.~D. Schwartz, {\it {$S$ -Matrix for massless particles}},
   \href{http://dx.doi.org/10.1103/PhysRevD.101.105001}{{\em Phys. Rev. D}
  {\bfseries 101} (2020) 105001}
  [\href{http://arxiv.org/abs/1911.06821}{{\ttfamily arXiv:1911.06821}}]
  [\href{http://inspirehep.net/search?p=find+Hannesdottir:2019opa}{{\ttfamily
  InSPIRE}}].

\bibitem{Hannesdottir:2019umk}
H.~Hannesdottir and M.~D. Schwartz, {\it {Finite $S$ matrix}},
  \href{http://dx.doi.org/10.1103/PhysRevD.107.L021701}{{\em Phys. Rev. D}
  {\bfseries 107} (2023) L021701}
  [\href{http://arxiv.org/abs/1906.03271}{{\ttfamily arXiv:1906.03271}}]
  [\href{http://inspirehep.net/search?p=find+Hannesdottir:2019umk}{{\ttfamily
  InSPIRE}}].

\bibitem{DelDuca:2001gu}
V.~Del~Duca and E.~W.~N. Glover, {\it {The High-energy limit of QCD at two
  loops}},  \href{http://dx.doi.org/10.1088/1126-6708/2001/10/035}{{\em JHEP}
  {\bfseries 10} (2001) 035}
  [\href{http://arxiv.org/abs/hep-ph/0109028}{{\ttfamily hep-ph/0109028}}]
  [\href{http://inspirehep.net/search?p=find+DelDuca:2001gu}{{\ttfamily
  InSPIRE}}].

\bibitem{DelDuca:2013ara}
V.~Del~Duca, G.~Falcioni, L.~Magnea, and L.~Vernazza, {\it {High-energy QCD
  amplitudes at two loops and beyond}},
  \href{http://dx.doi.org/10.1016/j.physletb.2014.03.033}{{\em Phys. Lett. B}
  {\bfseries 732} (2014) 233--240}
  [\href{http://arxiv.org/abs/1311.0304}{{\ttfamily arXiv:1311.0304}}]
  [\href{http://inspirehep.net/search?p=find+DelDuca:2013ara}{{\ttfamily
  InSPIRE}}].

\bibitem{DelDuca:2014cya}
V.~Del~Duca, G.~Falcioni, L.~Magnea, and L.~Vernazza, {\it {Analyzing
  high-energy factorization beyond next-to-leading logarithmic accuracy}},
  \href{http://dx.doi.org/10.1007/JHEP02(2015)029}{{\em JHEP} {\bfseries 02}
  (2015) 029} [\href{http://arxiv.org/abs/1409.8330}{{\ttfamily
  arXiv:1409.8330}}]
  [\href{http://inspirehep.net/search?p=find+DelDuca:2014cya}{{\ttfamily
  InSPIRE}}].

\bibitem{DelDuca:2011ae}
V.~Del~Duca, C.~Duhr, E.~Gardi, L.~Magnea, and C.~D. White, {\it {The Infrared
  structure of gauge theory amplitudes in the high-energy limit}},
  \href{http://dx.doi.org/10.1007/JHEP12(2011)021}{{\em JHEP} {\bfseries 12}
  (2011) 021} [\href{http://arxiv.org/abs/1109.3581}{{\ttfamily
  arXiv:1109.3581}}]
  [\href{http://inspirehep.net/search?p=find+DelDuca:2011ae}{{\ttfamily
  InSPIRE}}].

\bibitem{Pancheri:2016yel}
G.~Pancheri and Y.~N. Srivastava, {\it {Introduction to the physics of the
  total cross-section at LHC}: {A Review of Data and Models}},
  \href{http://dx.doi.org/10.1140/epjc/s10052-016-4585-8}{{\em Eur. Phys. J. C}
  {\bfseries 77} (2017) 150} [\href{http://arxiv.org/abs/1610.10038}{{\ttfamily
  arXiv:1610.10038}}]
  [\href{http://inspirehep.net/search?p=find+Pancheri:2016yel}{{\ttfamily
  InSPIRE}}].

\bibitem{Lipatov:1976zz}
L.~N. Lipatov, {\it {Reggeization of the Vector Meson and the Vacuum
  Singularity in Nonabelian Gauge Theories}},  {\em Sov. J. Nucl. Phys.}
  {\bfseries 23} (1976) 338--345
  [\href{http://inspirehep.net/search?p=find+Lipatov:1976zz}{{\ttfamily
  InSPIRE}}].

\bibitem{Kuraev:1976ge}
E.~A. Kuraev, L.~N. Lipatov, and V.~S. Fadin, {\it {Multi - Reggeon Processes
  in the Yang-Mills Theory}},  {\em Sov. Phys. JETP} {\bfseries 44} (1976)
  443--450 [\href{http://inspirehep.net/search?p=find+Kuraev:1976ge}{{\ttfamily
  InSPIRE}}].

\bibitem{Fadin:1975cb}
V.~S. Fadin, E.~A. Kuraev, and L.~N. Lipatov, {\it {On the Pomeranchuk
  Singularity in Asymptotically Free Theories}},
  \href{http://dx.doi.org/10.1016/0370-2693(75)90524-9}{{\em Phys. Lett. B}
  {\bfseries 60} (1975) 50--52}
  [\href{http://inspirehep.net/search?p=find+Fadin:1975cb}{{\ttfamily
  InSPIRE}}].

\bibitem{Balitsky:1978ic}
I.~I. Balitsky and L.~N. Lipatov, {\it {The Pomeranchuk Singularity in Quantum
  Chromodynamics}},  {\em Sov. J. Nucl. Phys.} {\bfseries 28} (1978) 822--829
  [\href{http://inspirehep.net/search?p=find+Balitsky:1978ic}{{\ttfamily
  InSPIRE}}].

\bibitem{Gorshkov:1966qd}
V.~G. Gorshkov, V.~N. Gribov, L.~N. Lipatov, and G.~V. Frolov, {\it {Double
  logarithmic asymptotics of quantum electrodynamics}},
  \href{http://dx.doi.org/10.1016/0031-9163(66)90701-3}{{\em Phys. Lett.}
  {\bfseries 22} (1966) 671--673}
  [\href{http://inspirehep.net/search?p=find+Gorshkov:1966qd}{{\ttfamily
  InSPIRE}}].

\bibitem{Gorshkov:1966ht}
V.~G. Gorshkov, V.~N. Gribov, L.~N. Lipatov, and G.~V. Frolov, {\it {Doubly
  logarithmic asymptotic behavior in quantum electrodynamics}},  {\em Yad.
  Fiz.} {\bfseries 6} (1967) 129
  [\href{http://inspirehep.net/search?p=find+Gorshkov:1966ht}{{\ttfamily
  InSPIRE}}].

\bibitem{Gorshkov:1966hu}
V.~G. Gorshkov, V.~N. Gribov, L.~N. Lipatov, and G.~V. Frolov, {\it {Backward
  electron - positron scattering at high-energies}},  {\em Yad. Fiz.}
  {\bfseries 6} (1967) 361
  [\href{http://inspirehep.net/search?p=find+Gorshkov:1966hu}{{\ttfamily
  InSPIRE}}].

\bibitem{Hahn:2000kx}
T.~Hahn, {\it {Generating Feynman diagrams and amplitudes with FeynArts 3}},
  \href{http://dx.doi.org/10.1016/S0010-4655(01)00290-9}{{\em Comput. Phys.
  Commun.} {\bfseries 140} (2001) 418--431}
  [\href{http://arxiv.org/abs/hep-ph/0012260}{{\ttfamily hep-ph/0012260}}]
  [\href{http://inspirehep.net/search?p=find+Hahn:2000kx}{{\ttfamily
  InSPIRE}}].

\bibitem{calcloop}
\url{https://e.gitee.com/multiloop-pku/repos/multiloop-pku/calcloop/sources}.

\bibitem{Anastasiou:2002yz}
C.~Anastasiou and K.~Melnikov, {\it {Higgs boson production at hadron colliders
  in NNLO QCD}},  \href{http://dx.doi.org/10.1016/S0550-3213(02)00837-4}{{\em
  Nucl. Phys. B} {\bfseries 646} (2002) 220--256}
  [\href{http://arxiv.org/abs/hep-ph/0207004}{{\ttfamily hep-ph/0207004}}]
  [\href{http://inspirehep.net/search?p=find+Anastasiou:2002yz}{{\ttfamily
  InSPIRE}}].

\bibitem{Anastasiou:2002qz}
C.~Anastasiou, L.~J. Dixon, and K.~Melnikov, {\it {NLO Higgs boson rapidity
  distributions at hadron colliders}},
  \href{http://dx.doi.org/10.1016/S0920-5632(03)80168-8}{{\em Nucl. Phys. B
  Proc. Suppl.} {\bfseries 116} (2003) 193--197}
  [\href{http://arxiv.org/abs/hep-ph/0211141}{{\ttfamily hep-ph/0211141}}]
  [\href{http://inspirehep.net/search?p=find+Anastasiou:2002qz}{{\ttfamily
  InSPIRE}}].

\bibitem{Anastasiou:2003yy}
C.~Anastasiou, L.~J. Dixon, K.~Melnikov, and F.~Petriello, {\it {Dilepton
  rapidity distribution in the Drell-Yan process at NNLO in QCD}},
  \href{http://dx.doi.org/10.1103/PhysRevLett.91.182002}{{\em Phys. Rev. Lett.}
  {\bfseries 91} (2003) 182002}
  [\href{http://arxiv.org/abs/hep-ph/0306192}{{\ttfamily hep-ph/0306192}}]
  [\href{http://inspirehep.net/search?p=find+Anastasiou:2003yy}{{\ttfamily
  InSPIRE}}].

\bibitem{Chetyrkin:1981qh}
K.~G. Chetyrkin and F.~V. Tkachov, {\it {Integration by parts: The algorithm to
  calculate $\beta$-functions in 4 loops}},
  \href{http://dx.doi.org/10.1016/0550-3213(81)90199-1}{{\em Nucl. Phys. B}
  {\bfseries 192} (1981) 159--204}
  [\href{http://inspirehep.net/search?p=find+Chetyrkin:1981qh}{{\ttfamily
  InSPIRE}}].

\bibitem{Laporta:2000dsw}
S.~Laporta, {\it {High-precision calculation of multiloop Feynman integrals by
  difference equations}},
  \href{http://dx.doi.org/10.1142/S0217751X00002159}{{\em Int. J. Mod. Phys. A}
  {\bfseries 15} (2000) 5087--5159}
  [\href{http://arxiv.org/abs/hep-ph/0102033}{{\ttfamily hep-ph/0102033}}]
  [\href{http://inspirehep.net/search?p=find+Laporta:2000dsw}{{\ttfamily
  InSPIRE}}].

\bibitem{Guan:2024byi}
X.~Guan, X.~Liu, Y.-Q. Ma, and W.-H. Wu, {\it {Blade: A package for
  block-triangular form improved Feynman integrals decomposition}},
  [\href{http://arxiv.org/abs/2405.14621}{{\ttfamily arXiv:2405.14621}}]
  [\href{http://inspirehep.net/search?p=find+Guan:2024byi}{{\ttfamily
  InSPIRE}}].

\bibitem{Liu:2018dmc}
X.~Liu and Y.-Q. Ma, {\it {Determining arbitrary Feynman integrals by vacuum
  integrals}},  \href{http://dx.doi.org/10.1103/PhysRevD.99.071501}{{\em Phys.
  Rev. D} {\bfseries 99} (2019) 071501}
  [\href{http://arxiv.org/abs/1801.10523}{{\ttfamily arXiv:1801.10523}}]
  [\href{http://inspirehep.net/search?p=find+Liu:2018dmc}{{\ttfamily
  InSPIRE}}].

\bibitem{Guan:2019bcx}
X.~Guan, X.~Liu, and Y.-Q. Ma, {\it {Complete reduction of integrals in
  two-loop five-light-parton scattering amplitudes}},
  \href{http://dx.doi.org/10.1088/1674-1137/44/9/093106}{{\em Chin. Phys. C}
  {\bfseries 44} (2020) 093106}
  [\href{http://arxiv.org/abs/1912.09294}{{\ttfamily arXiv:1912.09294}}]
  [\href{http://inspirehep.net/search?p=find+Guan:2019bcx}{{\ttfamily
  InSPIRE}}].

\bibitem{Peraro:2019svx}
T.~Peraro, {\it {$\text{FiniteFlow}$: multivariate functional reconstruction
  using finite fields and dataflow graphs}},
  \href{http://dx.doi.org/10.1007/JHEP07(2019)031}{{\em JHEP} {\bfseries 07}
  (2019) 031} [\href{http://arxiv.org/abs/1905.08019}{{\ttfamily
  arXiv:1905.08019}}]
  [\href{http://inspirehep.net/search?p=find+Peraro:2019svx}{{\ttfamily
  InSPIRE}}].

\bibitem{Kotikov:1990kg}
A.~V. Kotikov, {\it {Differential equations method: New technique for massive
  Feynman diagrams calculation}},
  \href{http://dx.doi.org/10.1016/0370-2693(91)90413-K}{{\em Phys. Lett. B}
  {\bfseries 254} (1991) 158--164}
  [\href{http://inspirehep.net/search?p=find+Kotikov:1990kg}{{\ttfamily
  InSPIRE}}].

\bibitem{Liu:2022chg}
X.~Liu and Y.-Q. Ma, {\it {AMFlow: A Mathematica package for Feynman integrals
  computation via auxiliary mass flow}},
  \href{http://dx.doi.org/10.1016/j.cpc.2022.108565}{{\em Comput. Phys.
  Commun.} {\bfseries 283} (2023) 108565}
  [\href{http://arxiv.org/abs/2201.11669}{{\ttfamily arXiv:2201.11669}}]
  [\href{http://inspirehep.net/search?p=find+Liu:2022chg}{{\ttfamily
  InSPIRE}}].

\bibitem{Chen:2022vzo}
X.~Chen, X.~Guan, C.-Q. He, X.~Liu, and Y.-Q. Ma, {\it {Heavy-Quark Pair
  Production at Lepton Colliders at NNNLO in QCD}},
  \href{http://dx.doi.org/10.1103/PhysRevLett.132.101901}{{\em Phys. Rev.
  Lett.} {\bfseries 132} (2024) 101901}
  [\href{http://arxiv.org/abs/2209.14259}{{\ttfamily arXiv:2209.14259}}]
  [\href{http://inspirehep.net/search?p=find+Chen:2022vzo}{{\ttfamily
  InSPIRE}}].

\bibitem{Bi:2023bnq}
H.-Y. Bi, L.-H. Huang, R.-J. Huang, Y.-Q. Ma, and H.-M. Yu, {\it {Electroweak
  Corrections to Double Higgs Production at the LHC}},
  \href{http://dx.doi.org/10.22323/1.478.0120}{{\em Phys. Rev. Lett.}
  {\bfseries 132} (2024) 231802}
  [\href{http://arxiv.org/abs/2311.16963}{{\ttfamily arXiv:2311.16963}}]
  [\href{http://inspirehep.net/search?p=find+Bi:2023bnq}{{\ttfamily InSPIRE}}].

\bibitem{Liu:2017jxz}
X.~Liu, Y.-Q. Ma, and C.-Y. Wang, {\it {A Systematic and Efficient Method to
  Compute Multi-loop Master Integrals}},
  \href{http://dx.doi.org/10.1016/j.physletb.2018.02.026}{{\em Phys. Lett. B}
  {\bfseries 779} (2018) 353--357}
  [\href{http://arxiv.org/abs/1711.09572}{{\ttfamily arXiv:1711.09572}}]
  [\href{http://inspirehep.net/search?p=find+Liu:2017jxz}{{\ttfamily
  InSPIRE}}].

\bibitem{Liu:2020kpc}
X.~Liu, Y.-Q. Ma, W.~Tao, and P.~Zhang, {\it {Calculation of Feynman loop
  integration and phase-space integration via auxiliary mass flow}},
  \href{http://dx.doi.org/10.1088/1674-1137/abc538}{{\em Chin. Phys. C}
  {\bfseries 45} (2021) 013115}
  [\href{http://arxiv.org/abs/2009.07987}{{\ttfamily arXiv:2009.07987}}]
  [\href{http://inspirehep.net/search?p=find+Liu:2020kpc}{{\ttfamily
  InSPIRE}}].

\bibitem{Liu:2021wks}
X.~Liu and Y.-Q. Ma, {\it {Multiloop corrections for collider processes using
  auxiliary mass flow}},
  \href{http://dx.doi.org/10.1103/PhysRevD.105.L051503}{{\em Phys. Rev. D}
  {\bfseries 105} (2022) L051503}
  [\href{http://arxiv.org/abs/2107.01864}{{\ttfamily arXiv:2107.01864}}]
  [\href{http://inspirehep.net/search?p=find+Liu:2021wks}{{\ttfamily
  InSPIRE}}].

\bibitem{Liu:2022mfb}
Z.-F. Liu and Y.-Q. Ma, {\it {Determining Feynman Integrals with Only Input
  from Linear Algebra}},
  \href{http://dx.doi.org/10.1103/PhysRevLett.129.222001}{{\em Phys. Rev.
  Lett.} {\bfseries 129} (2022) 222001}
  [\href{http://arxiv.org/abs/2201.11637}{{\ttfamily arXiv:2201.11637}}]
  [\href{http://inspirehep.net/search?p=find+Liu:2022mfb}{{\ttfamily
  InSPIRE}}].

\bibitem{Caffo:2008aw}
M.~Caffo, H.~Czyz, M.~Gunia, and E.~Remiddi, {\it {BOKASUN: A Fast and precise
  numerical program to calculate the Master Integrals of the two-loop sunrise
  diagrams}},  \href{http://dx.doi.org/10.1016/j.cpc.2008.10.011}{{\em Comput.
  Phys. Commun.} {\bfseries 180} (2009) 427--430}
  [\href{http://arxiv.org/abs/0807.1959}{{\ttfamily arXiv:0807.1959}}]
  [\href{http://inspirehep.net/search?p=find+Caffo:2008aw}{{\ttfamily
  InSPIRE}}].

\bibitem{Czakon:2008zk}
M.~Czakon, {\it {Tops from Light Quarks: Full Mass Dependence at Two-Loops in
  QCD}},  \href{http://dx.doi.org/10.1016/j.physletb.2008.05.028}{{\em Phys.
  Lett. B} {\bfseries 664} (2008) 307--314}
  [\href{http://arxiv.org/abs/0803.1400}{{\ttfamily arXiv:0803.1400}}]
  [\href{http://inspirehep.net/search?p=find+Czakon:2008zk}{{\ttfamily
  InSPIRE}}].

\bibitem{Broadhurst:1991fy}
D.~J. Broadhurst, N.~Gray, and K.~Schilcher, {\it {Gauge invariant on-shell
  Z(2) in QED, QCD and the effective field theory of a static quark}},
  \href{http://dx.doi.org/10.1007/BF01412333}{{\em Z. Phys. C} {\bfseries 52}
  (1991) 111--122}
  [\href{http://inspirehep.net/search?p=find+Broadhurst:1991fy}{{\ttfamily
  InSPIRE}}].

\bibitem{Melnikov:2000zc}
K.~Melnikov and T.~van Ritbergen, {\it {The Three loop on-shell renormalization
  of QCD and QED}},
  \href{http://dx.doi.org/10.1016/S0550-3213(00)00526-5}{{\em Nucl. Phys. B}
  {\bfseries 591} (2000) 515--546}
  [\href{http://arxiv.org/abs/hep-ph/0005131}{{\ttfamily hep-ph/0005131}}]
  [\href{http://inspirehep.net/search?p=find+Melnikov:2000zc}{{\ttfamily
  InSPIRE}}].

\bibitem{Thirring:1950cy}
W.~E. Thirring, {\it {Radiative corrections in the nonrelativistic limit}},
  {\em Phil. Mag. Ser. 7} {\bfseries 41} (1950) 1193--1194
  [\href{http://inspirehep.net/search?p=find+Thirring:1950cy}{{\ttfamily
  InSPIRE}}].

\bibitem{Dittmaier:1997dx}
S.~Dittmaier, {\it {Thirring's low-energy theorem and its generalizations in
  the electroweak standard model}},
  \href{http://dx.doi.org/10.1016/S0370-2693(97)00888-5}{{\em Phys. Lett. B}
  {\bfseries 409} (1997) 509--516}
  [\href{http://arxiv.org/abs/hep-ph/9704368}{{\ttfamily hep-ph/9704368}}]
  [\href{http://inspirehep.net/search?p=find+Dittmaier:1997dx}{{\ttfamily
  InSPIRE}}].

\bibitem{Smirnov:1997gx}
V.~A. Smirnov, {\it {Asymptotic expansions of two loop Feynman diagrams in the
  Sudakov limit}},  \href{http://dx.doi.org/10.1016/S0370-2693(97)00545-5}{{\em
  Phys. Lett. B} {\bfseries 404} (1997) 101--107}
  [\href{http://arxiv.org/abs/hep-ph/9703357}{{\ttfamily hep-ph/9703357}}]
  [\href{http://inspirehep.net/search?p=find+Smirnov:1997gx}{{\ttfamily
  InSPIRE}}].

\bibitem{Beneke:1997zp}
M.~Beneke and V.~A. Smirnov, {\it {Asymptotic expansion of Feynman integrals
  near threshold}},
  \href{http://dx.doi.org/10.1016/S0550-3213(98)00138-2}{{\em Nucl. Phys. B}
  {\bfseries 522} (1998) 321--344}
  [\href{http://arxiv.org/abs/hep-ph/9711391}{{\ttfamily hep-ph/9711391}}]
  [\href{http://inspirehep.net/search?p=find+Beneke:1997zp}{{\ttfamily
  InSPIRE}}].

\bibitem{Hou:2025ovb}
J.-Y. Hou, J.~Wang, and D.-J. Zhang, {\it {Region analysis of H
  {\textrightarrow} {\ensuremath{\gamma}}{\ensuremath{\gamma}} via a bottom
  quark loop}},  \href{http://dx.doi.org/10.1007/JHEP06(2025)164}{{\em JHEP}
  {\bfseries 06} (2025) 164} [\href{http://arxiv.org/abs/2501.11824}{{\ttfamily
  arXiv:2501.11824}}]
  [\href{http://inspirehep.net/search?p=find+Hou:2025ovb}{{\ttfamily
  InSPIRE}}].

\bibitem{Bell:2022ott}
G.~Bell, P.~B{\"o}er, and T.~Feldmann, {\it {Muon-electron backward scattering:
  a prime example for endpoint singularities in SCET}},
  \href{http://dx.doi.org/10.1007/JHEP09(2022)183}{{\em JHEP} {\bfseries 09}
  (2022) 183} [\href{http://arxiv.org/abs/2205.06021}{{\ttfamily
  arXiv:2205.06021}}]
  [\href{http://inspirehep.net/search?p=find+Bell:2022ott}{{\ttfamily
  InSPIRE}}].

\bibitem{Donoghue:2014mpa}
J.~F. Donoghue, B.~K. El-Menoufi, and G.~Ovanesyan, {\it {Regge behavior in
  effective field theory}},
  \href{http://dx.doi.org/10.1103/PhysRevD.90.096009}{{\em Phys. Rev. D}
  {\bfseries 90} (2014) 096009}
  [\href{http://arxiv.org/abs/1405.1731}{{\ttfamily arXiv:1405.1731}}]
  [\href{http://inspirehep.net/search?p=find+Donoghue:2014mpa}{{\ttfamily
  InSPIRE}}].

\bibitem{PrimEx:2019zre}
{\bfseries PrimEx} , P.~Ambrozewicz {\em et al.}, {\it {High Precision
  Measurement of Compton Scattering in the 5 GeV region}},
  \href{http://dx.doi.org/10.1016/j.physletb.2019.134884}{{\em Phys. Lett. B}
  {\bfseries 797} (2019) 134884}
  [\href{http://arxiv.org/abs/1903.05529}{{\ttfamily arXiv:1903.05529}}]
  [\href{http://inspirehep.net/search?p=find+PrimEx:2019zre}{{\ttfamily
  InSPIRE}}].

\bibitem{GlueX:2025hve}
{\bfseries GlueX} , F.~Afzal {\em et al.}, {\it {Measurement of the total
  compton scattering cross section between 6.5 and 11 GeV}},
  \href{http://dx.doi.org/10.1016/j.physletb.2025.139914}{{\em Phys. Lett. B}
  {\bfseries 870} (2025) 139914}
  [\href{http://arxiv.org/abs/2505.07994}{{\ttfamily arXiv:2505.07994}}]
  [\href{http://inspirehep.net/search?p=find+GlueX:2025hve}{{\ttfamily
  InSPIRE}}].

\bibitem{doi:10.1126/science.aap7706}
R.~H. Parker, C.~Yu, W.~Zhong, B.~Estey, and H.~Müller, {\it Measurement of
  the fine-structure constant as a test of the standard model},
  \href{http://dx.doi.org/10.1126/science.aap7706}{{\em Science} {\bfseries
  360} (2018) 191--195}
  [\href{http://arxiv.org/abs/https://www.science.org/doi/pdf/10.1126/science.aap7706}{{\ttfamily
  https://www.science.org/doi/pdf/10.1126/science.aap7706}}]
  [\href{http://inspirehep.net/search?p=find+doi:10.1126/science.aap7706}{{\ttfamily
  InSPIRE}}]. \url{https://www.science.org/doi/abs/10.1126/science.aap7706}.

\bibitem{sturm2014high}
S.~Sturm, F.~K{\"o}hler, J.~Zatorski, A.~Wagner, Z.~Harman, G.~Werth, W.~Quint,
  C.~H. Keitel, and K.~Blaum, {\it High-precision measurement of the atomic
  mass of the electron},  {\em Nature} {\bfseries 506} (2014) 467--470
  [\href{http://inspirehep.net/search?p=find+sturm2014high}{{\ttfamily
  InSPIRE}}].

\bibitem{Ellis:2016jkw}
J.~Ellis, {\it {TikZ-Feynman: Feynman diagrams with TikZ}},
  \href{http://dx.doi.org/10.1016/j.cpc.2016.08.019}{{\em Comput. Phys.
  Commun.} {\bfseries 210} (2017) 103--123}
  [\href{http://arxiv.org/abs/1601.05437}{{\ttfamily arXiv:1601.05437}}]
  [\href{http://inspirehep.net/search?p=find+Ellis:2016jkw}{{\ttfamily
  InSPIRE}}].

\end{thebibliography}\endgroup

\end{document}